\documentclass[aip,[preprint]{revtex4-1}
\usepackage[margin=1in, paperwidth=8.5in, paperheight=11in]{geometry}
\usepackage{amsmath,amssymb}
\usepackage{setspace}
\usepackage{graphicx}
\newcommand{\p}[2]{\frac{\partial #1}{\partial #2}}

\newcommand{\dnq}[3]{\frac{d^{#3} #1}{d #2^{#3}}}
\newcommand{\de}[2]{\frac{d #1}{d #2}}
\newcommand{\R}{\mathbb{R}}
\def\cross{\times}
\def\calb{\mathcal{B}}

\def\calm{\mathcal{M}}

\begin{document}
\title{Travelling Waves in Hall-MHD and the Ion-Acoustic Shock Structure}
\author{George I. Hagstrom}
\affiliation{Courant Institute of Mathematical Sciences}
 \author{Eliezer Hameiri}
 \affiliation{Courant Institute of Mathematical Sciences}

\begin{abstract} Hall-MHD is a mixed hyperbolic-parabolic partial differential equation that describes the dynamics of an ideal two fluid plasma with massless electrons. We study the only shock wave family that exists in this system (the other discontinuities being contact discontinuities and not shocks). We study planar travelling wave solutions and we find solutions with discontinuities in the hydrodynamic variables, which arise due to the presence of real characteristics in Hall-MHD. We introduce a small viscosity into the equations and use the method of matched asymptotic expansions to show that solutions with a discontinuity satisfying the Rankine-Hugoniot conditions and also an entropy condition have continuous shock structures. The lowest order inner equations reduce to the compressible Navier-Stokes equations, plus an equation which implies the constancy of the magnetic field inside the shock structure. We are able to show that the current is discontinuous across the shock, even as the magnetic field is continuous, and that the lowest order outer equations, which are the equations for travelling waves in inviscid Hall-MHD, are exactly integrable. We show that the inner and outer solutions match, which allows us to construct a family of uniformly valid continuous composite solutions that become discontinuous when the diffusivity vanishes.

\end{abstract}

\maketitle

\section{Introduction}

% Note to self: Edit this thoroughly after composition is complete!

Ideal magnetohydrodynamics (MHD) is one of the most important models that describes a plasma and is a common starting point for theoretical descriptions of plasma phenomena. As plasma physics has developed, phenomena that are beyond the reach of MHD have seen increased attention. This trend has accelerated in recent years, and a number of extended-MHD theories have
become important topics of research.
Hall-magnetohydrodynamics (Hall-MHD) is an extended-MHD model in between two-fluid 
theory and MHD. Hall-MHD describes the behavior of a plasma at length scales comparable with the ion skin-depth and time scales comparable to the ion-cyclotron frequency, but is simpler than two-fluid theory because it has fewer variables.  
Hall-MHD has helped researchers progress towards the solution of a number of difficult problems, particularly magnetic reconnection in plasmas with very low resistivity. Spontaneous reconnection leads to violent phenomena: solar flares in the corona, magnetic substorms in magnetospheric plasmas, and sawtooth oscillations in tokamaks. These effects are too strong and rapid to be explained using resistive-MHD, and in each of these cases Hall-MHD has shown promise\cite{BhattMaWang, WangBhattMa}. Hall-MHD has been used as an enhanced description of other phenomena that involve small length scales, for instance the dissipative range of turbulence\cite{MahajanKrishnan} and ballooning modes\cite{HameiriHallWaves}.
The Hall effect enables the rapid penetration of a magnetic field into a plasma inside plasma opening switches, which are crucial components of pulsed plasma devices, radiation sources, and charged particle beams\cite{HallPOS}. Hall effect based thrusters provide high efficiency propulsion to spacecraft and satellites\cite{HallThrusters}. Further development of Hall-MHD is motivated both by a desire to understand difficult physical problems and
to develop new technologies.

Hall-MHD is a conservative two-fluid system in which the electron fluid has no inertia. Ions and electrons in a plasma begin to decouple due to the gyro-motion of the ions, on length scales comparable to the ion-skin depth. This gives rise to the Hall effect, which is the generation of an electric field proportional to ${\bf J}\cross{\bf B}$ (where ${\bf J}$ is the current density and ${\bf B}$ is the magnetic field) and which is included along with the electron pressure gradient in the generalized Ohm's law.
These terms are a singular perturbation of MHD, giving rise to second order derivatives of the magnetic field and electron pressure. Unlike MHD, Hall-MHD is a mixed hyperbolic-parabolic system (also referred to as parabolic-degenerate\cite{TataronisRosenau}). The parabolicity comes from the presence of dispersive terms, and waves in Hall-MHD exhibit both hyperbolic and dispersive properties\cite{HameiriHallWaves}. 

Here we develop the theory of a one-dimensional shock wave in Hall-MHD, for the particular case of isothermal electrons. Nonlinear hyperbolic systems can form shocks, and MHD is an example of such a system; it
has a rich family of shock waves and it was one of the first systems for which there was a detailed study of shocks\cite{Anderson,Smoller}. The appearance of shock waves in MHD foreshadowed their importance in plasma physics. Hall-MHD is different due to effects of dispersion, which may alter shock formation and complicate the use of the standard methods for studying shock waves in hyperbolic systems. Because Hall-MHD is a mixed hyperbolic-parabolic system, it is sensible to expect these standard methods to be useful for finding shocks associated with the real characteristics, which is the course taken here. There have been some previous works on shock waves in 
Hall-MHD. This work differs from these studies mostly because it looks at the vanishing viscosity limit of viscous solutions, though we also use a more general equation of state. Previous work dealt only with polytropic ions and \emph{zero} electron pressure. In that case, jump conditions for shocks were written down by Leonard\cite{Leonard} in 1966 and later by Tataronis and Rosenau\cite{TataronisRosenau} and Lee and Wu\cite{LeeWuHall}. 
Shocks with a rotational discontinuity in the magnetic field have been shown to form in axisymmetric geometries, leading to KMC (Kingsep, Mokhov, Chukbar) waves\cite{KMC,Degond,DreherRubanGrauer}, which have been observed in plasma opening switches\cite{HallPOS}. In the non-planar case, using the full set of Hall-MHD equations, more general jump conditions, which apply to more than just the ion-acoustic characteristic family, were developed by Hameiri\cite{HameiriHall}, which was the motivation for this project.

The planar shock problem in Hall-MHD is a special case of the more general problem of finding planar shock waves in mixed hyperbolic-parabolic systems. The simplest and perhaps first known example of a mixed hyperbolic-parabolic system with shocks is the compressible Euler equation augmented with non-zero heat conductivity. It was first shown by Rayleigh\cite{Rayleigh} that the presence of heat conduction alone is insufficient to guarantee smoothness of solutions to the equations. The exposition in Gilbarg\cite{Gilbarg} contains a rigorous treatment of this case and a proof of the existence of shock structures for discontinuous solutions of that model that satisfy an entropy condition. A framework for the general problem was developed by Germain\cite{GermainEnglish} in the context of subshocks, which occur when there are multiple dissipative mechanisms with order of magnitude differences between their strength. 

By assuming that the magnetic field ${\bf B}$ is continuous we are able to recover the jump conditions for the ion-acoustic shock (called such because it is associated with a characteristic family that propagates at the sound speed in the plasma) from the general conditions in Hameiri\cite{HameiriHall}. The jump conditions include those of an ideal gas, but there is also a jump in the derivative of the magnetic field across the shock. Therefore the shock waves are an analog of the shock waves that develop in the compressible Euler equations of fluid dynamics, but with an additional condition involving the current, and are associated with characteristics that propagate information at the sound speed of the plasma.  We study the shock structure using the viscosity method\cite{Smoller}, which involves adding viscosity to the equations and calculating a family of one-dimensional travelling wave solutions, parametrized by the viscosity, that become discontinuous in the limit that the viscosity vanishes. In one spatial dimension plus time ($z,t$ coordinates), travelling waves depend only on a single variable $z'=z-Ut$, which reduces the problem of calculating shock structures to a finite degree-of-freedom dynamical systems problem. Because the inviscid equations are second order, the shock structures are of a different type than shock structures in MHD and other hyperbolic systems.
The states on opposite sides of the shock are not stationary points of the travelling wave system, which means that generalized solutions of the inviscid system are non-constant away from the shock, which is different from what is usually seen in the hyperbolic case, and also necessary due to the fact that Hall-MHD is a mixed hyperbolic-parabolic system. 

In the next section we will introduce the equations of Hall-MHD and describe their basic mathematical properties. We will then
add small viscosity and heat conductivity and reduce the resulting equations to ordinary differential equations (ODEs) for one-dimensional travelling waves. We will note that the viscous terms are a singular perturbation, and
use the method of matched asymptotic expansions to derive inner and outer equations, the inner equations describing the behavior of the dynamical variables within a shock. We will show that the values of the inner solution at $\pm\infty$, in the coordinates describing the shock structure, must satisfy the jump conditions for the ion-acoustic
shock, and that for any pair of states that satisfy these jump conditions and an entropy condition, there is an asymptotic solution that depends on the diffusivity that
has the appropriate discontinuity in the limit of vanishing diffusivity. We will show that the inner and outer solutions match, from which we will conclude that the ion-acoustic shocks in Hall-MHD have shock structures.

\section{Hall-MHD Equations}

Hall-MHD is a system of second-order quasilinear partial-differential equations, which we give below:

\begin{subequations}\begin{gather}
\p{\rho}{t}+\nabla\cdot(\rho {\bf u}) =0 \label{eq:HallRho}\\
\p{(\rho{\bf u})}{t}+\nabla\cdot\left(\rho{\bf uu}+(p+\frac{1}{2}{\bf B}^2){\bf I}-{\bf BB}\right) =0 \\
\p{{\bf B}}{t}+\nabla\cross {\bf E} =0 \label{eq:flux}\\
\p{S_i}{t}+{\bf u}\cdot\nabla S_i = 0 \label{eq:SIon} \\
{\bf E}+{\bf u}\cross {\bf B} = \frac{\epsilon}{\rho}\left({\bf J}\cross{\bf B}-\nabla p_e\right) \label{eq:Ohm} \\
\nabla\cross{\bf B} = {\bf J} \\
\nabla\cdot{\bf B} =0  \label{eq:HallDiv}
\end{gather}\end{subequations}

The density is $\rho$, the ion velocity is ${\bf u}=u_x\hat{{\bf x}}+u_y\hat{{\bf y}}+u_z\hat{{\bf z}}$, the total pressure $p=p_e+p_i$ is the sum of the pressures of the electrons and ions respectively, and the magnetic field is ${\bf B}=B_x\hat{{\bf x}}+B_y\hat{{\bf y}}+B_z\hat{{\bf z}}$. The ion mass-to-charge ratio is $\epsilon$, which expresses the magnitude of the Hall effect and which will not be assumed to be small in this paper. We assume that the ions are adiabatic, which implies the constancy of the specific ion entropy (which means entropy per unit mass) along streamlines, Eq \ref{eq:SIon}. The internal energy $e$ is the sum of the internal energy of the ion and electron fluids, $e=e_i+e_e$. 

The electron pressure gradient term makes Hall-MHD an explicitly two-fluid theory, and an additional equation of state for the electron fluid is necessary to close the set of equations. In this paper we choose isothermal electrons, $p_e=R T_e\rho$, where the electron temperature $T_e$ is a constant throughout the fluid and $R$ is a 
constant proportional to Boltzmann's constant. The internal energy of the isothermal electrons is given by 
$e_e=RT_e\log(\rho/\rho_0)$, where $\rho_0$ is some constant reference value for $\rho$.

Using 
Eqs. \ref{eq:HallRho} to \ref{eq:HallDiv} we derive the energy equation:

\begin{equation}
\p{}{t}\left(\frac{\rho {\bf u}^2}{2}+\rho e+\frac{1}{2}{\bf B}^2\right)
+\nabla\cdot\left(\left(\frac{\rho {\bf u}^2}{2}+\rho e +p\right){\bf u}+{\bf E\cross B}-\epsilon RT_e\log(\frac{\rho}{\rho_0}){\bf J}\right) =0 \label{eq:HallEnergy}\\
\end{equation}

Ohm's law, Eq. \ref{eq:Ohm}, is different than in MHD due to the appearance of ${\bf J}$ and $p_e$.
The current and electron pressure terms do not appear in MHD. On length scales comparable with the ion skin-depth $c/\omega_{p_i}$ the ion gyromotion causes a difference in the motion of electrons and ions, and the Hall current term becomes
comparable to the macroscopic velocity ${\bf u}$. Hall-MHD is an extension of MHD to these length scales; it captures some of the two fluid effects that MHD cannot. In the rest of the paper, we use Eq. \ref{eq:Ohm} to write the electric field in terms of the other dynamical variables.

\subsection{One-Dimensional Hall-MHD equations}

We assume that all dynamical variables are functions only of the coordinate $z$. We denote the decomposition of vectors into parallel and perpendicular directions with respect to planes of constant $z$ by ${\bf v}={\bf v}_{s}+{\bf \hat{z}}v_{z}$. We also introduce the matrix $\calm=\begin{pmatrix}
0 & -1 \\
1 & 0 \end{pmatrix}$, such that ${\bf J}_s=\calm \de{{\bf B}_s}{z}$ and $\calm^{-1}=-\calm$. Then the equations of Hall-MHD reduce to:

\begin{subequations}\begin{gather}
\p{\rho}{t}+\p{}{z}\left(\rho u_z\right) = 0 \label{eq:1DHallmassflux} \\
\p{}{t}\left(\rho {\bf u}\right)+\p{}{z}\left(\rho u_z {\bf u}+{\bf \hat{z}}(p+\frac{1}{2}B^2)-B_z{\bf B}\right)= 0 \label{eq:1DHallmomentumxflux} \\
\p{{\bf B}_{s}}{t}+\p{}{z}\left(u_z{\bf B}_{s}-B_z{\bf u}_{s}+\frac{\epsilon B_z}{\rho}\calm\de{{\bf B}_{s}}{z}\right) = 0 \label{eq:1DHallBxflx}\\
\p{B_z}{z} = 0 \label{eq:BzConstant}
\end{gather}
\begin{gather}
\p{}{t}\left(\frac{\rho {\bf u}^2}{2}+\rho e+\frac{1}{2}{\bf B}^2\right)+ 
\p{}{z}\left(\left[\frac{\rho{\bf u}^2}{2}+\rho e+p+B^2\right]u_z-{\bf B}\cdot{\bf u}B_z+\frac{\epsilon B_z{\bf B}_{s}}{\rho}\cdot\calm\p{{\bf B}_{s}}{z}\right) = 0 \label{eq:1DHallenergyflux} \\
\p{S_i}{t}+u_z\p{S_i}{z}=0 \label{eq:1DHallEntropy}
\end{gather}\end{subequations}

Note that the last term in Eq. \ref{eq:HallEnergy} disappears in one-dimension due to the vanishing of $J_z=\p{B_y}{x}-\p{B_x}{y}=0$ on account of one-dimensionality.
We have the choice of using Eq. \ref{eq:1DHallenergyflux} or Eq. \ref{eq:1DHallEntropy}; the two are not independent, and the important difference is that Eq. \ref{eq:1DHallenergyflux} is in conservation form and valid across discontinuities while Eq. \ref{eq:1DHallEntropy} is not valid there.

\subsection{Characteristics and Jump Conditions}

The Hall-MHD equations have two families of real characteristics with finite propagation speed\cite{TataronisRosenau}, one associated with a contact discontinuity and propagating at velocity $u_z$, and the other associated with the ion-acoustic wave and propagating at velocity $u_z\pm c_s$, where $c_s=\frac{\partial p}{\partial \rho}$ is the sound speed of the fluid, which has contributions from both the ion and electron pressure so that $c_s^2$ can be written $c_s^2=c_i^2+c_e^2$. The ion-acoustic characteristic is a direct analog of the sound wave in the Euler equation, suggesting that non-linear steepening can lead to the formation of shocks. 

The equations of Hall-MHD can be written in integral form as a system of conservation laws; the equations represent the conservation of mass, momentum, magnetic flux, and energy. There is no dissipation and the ion entropy is simply advected in regions where the dynamical variables are continuous.  We consider possibly discontinuous solutions of the system of conservation laws under the assumption that ${\bf B}$, but not necessarily its derivatives, are continuous. These assumptions isolate the ion-acoustic discontinuity\cite{TataronisRosenau}. In a reference frame where $u_z'$ measures velocity relative to the shock front, which moves at constant velocity $U{\bf \hat{z}}$ in the lab frame (so that $u_z'=u_z-U$),  we define the mass flux $m=\rho u_z'$ (which we assume to be nonzero). These assumptions yield the Rankine-Hugoniot conditions:

\begin{subequations}\begin{gather}
[m] = 0 \label{eq:massflux1} \\
m[u_z']+[p] = 0 \label{eq:momentumflux1} \\
[{\bf u}_{s}] = 0 \label{eq:transversemomentum1}\\
\left[u_z'{\bf B_{s}}+\frac{\epsilon B_z}{\rho}{\bf J_{s}}\right] = 0 \label{eq:magneticflux1} \\
\left[\left(\frac{1}{2}\rho {\bf u'}^2+\rho e+p\right)u_z'\right] = 0 \label{eq:energyflux1} \\
[T_e]=0  \label{eq:Tcontinuity}\\
[{\bf B}]=0 \label{eq:Bcontinuity1}
\end{gather}\end{subequations}
Here the symbol $[f]=\lim_{z\rightarrow z_0+}f(z)-\lim_{z\rightarrow z_0-}f(z)$ stands for the difference between the limits from the right and left of a function with at worst a jump discontinuity at $z=z_0$, the shock's location. 
Eq. \ref{eq:massflux1} to \ref{eq:transversemomentum1} and \ref{eq:energyflux1} are the equations for the classical hydrodynamic shock that occurs in the compressible Euler equation. Eq. \ref{eq:magneticflux1} gives the discontinuity in $\bf J$ in terms of the hydrodynamic variables, and by assumption the magnetic field remains continuous across the shock. Eq. \ref{eq:energyflux1} is derived from the integral of Eq. \ref{eq:1DHallenergyflux}, which on its own would produce:
\begin{equation}
\left[\left(\frac{1}{2}\rho {\bf u}^2+\rho e\right)u_z'+pu_z+{\bf B}^2u_z-{\bf B}\cdot {\bf u}B_z-\frac{m{\bf B_s^2}}{\rho}\right] = 0
\end{equation}
Using the other equations we can eliminate the dependence on $U$ so that only the relative velocity $u_z'$ appears,
recovering Eq. \ref{eq:energyflux1}.
 Later we show that when a small viscous term is included, there is a travelling wave solution that converges to a discontinuous function satisfying the chosen jump conditions for each possible jump that satisfies the entropy condition $m[S]>0$. Jump condition Eq. \ref{eq:Tcontinuity} holds for a general Hall-MHD discontinuity\cite{HameiriHall}, and one of the advantages of considering the electron fluid isothermal is that there is no need to assure that this condition holds.

\section{Hall-MHD Travelling Waves}

A piecewise continuous function that solves the equations of Hall-MHD wherever it is continuous, and which satisfies 
the Rankine-Hugoniot conditions, in one dimension Eqs. \ref{eq:massflux1} to \ref{eq:Bcontinuity1} across any of its discontinuities, is called a generalized solution of Hall-MHD. The differential equations of Hall-MHD, Eq. \ref{eq:1DHallmassflux} to \ref{eq:1DHallenergyflux} augmented with the Rankine-Hugoniot conditions, are insufficient to guarantee a unique solution
of the equations. One way to determine a unique, physically correct, solution is to incorporate the physics that happens inside the shock layer. The goal of this is to derive an entropy condition, which restricts the set of allowable discontinuities. One method for deriving an entropy condition is the viscosity method. The viscosity method involves the introduction of small dissipative terms in the equations (in fluid mechanics models these often involve the viscosity) and consideration of the limits of solutions as the viscosity coefficients are decreased to zero. 
We add viscosity and heat conductivity to regularize Hall-MHD and to generate smooth shock structures and search for travelling wave
solutions, reducing the partial differential equations into ordinary differential equations. The purpose of the viscosity method is to show that some discontinuous solutions to the inviscid problem are the limits of smooth solutions of the viscous equations. It is generally impossible to prove this for systems of equations, but for the special case of travelling wave solutions it is often feasible 
because the equations become ordinary differential equations\cite{Smoller}.  Travelling waves are solutions that depend on a single variable $z'=z-Ut$, to be referred to as $z$ from now on, and represent steady profiles travelling in the $\hat{\bf{z}}$ direction with velocity $U$. We use this ansatz on equations \ref{eq:1DHallmassflux} to \ref{eq:1DHallenergyflux}, redefining $u_z$
as $u_z-U$. To make the system viscous we add $\mu\dnq{{\bf u}}{z}{2}$ to the right hand side of Eq. \ref{eq:1DHallmomentumxflux} and $\de{}{z}\left(\mu\frac{1}{2}\de{}{z}|{\bf u}|^2+\kappa\de{T_i}{z}\right)$ to equation \ref{eq:1DHallenergyflux}. Here the kinematic viscosity $\mu$ and heat conductivity $\kappa$ are constants. Because all of the equations remain in divergence form after the addition of the viscous terms, we integrate each of them once, using the equations of magnetic flux to simplify the energy equation. The equations for travelling waves that result from this procedure are:

\begin{subequations}\label{eq:travellingwaves}
\begin{gather} 
\rho u_z = m \label{eq:twaves1} \\ 
\mu \de{{\bf u}}{z}=m {\bf u} +{\bf \hat{z}}\left(p+\frac{1}{2}B^2\right)- B_z {\bf B} - {\bf C}_H \label{eq:twaves2}\\
-\frac{\epsilon B_z}{\rho}\calm\de{{\bf B}_{s}}{z}=u_z {\bf B}_{s} -B_z{\bf u}_{s} - {\bf C}_B \\ 
\mu \frac{1}{2}\de{}{z} |{\bf u}|^2+\kappa\de{T_i}{z}=\frac{m}{2} |{\bf u}|^2 + \rho e u_z+ p u_z + {\bf B}_{s}\cdot{\bf C}_B -C_E \label{eq:twaves4} 
\end{gather}\end{subequations}

Without loss of generality, in the rest of this paper, it will be assumed that:
\begin{equation}
m>0.
\end{equation}

The choice of viscous terms can impact the types of shocks that have structure.
For hyperbolic PDEs shock structures correspond to families of travelling waves, each of which is a heteroclinic orbit that connects 
constant states (equlibrium points) of the travelling wave equations. The stationary points represent the state of the solution on opposite sides of the shock. The theory of these equations has been developed extensively and there are powerful topological methods for proving the existence of heteroclinic orbits\cite{Smoller}. These methods have successfully led to the solution of the shock structure problem in ideal-magnetohydrodynamics\cite{Smoller,ConleySmoller}.

The shock structure problem in Hall-MHD differs from that of a hyperbolic conservation law because the states on the opposite side of the shock are not stationary points. Despite this, it could be possible to treat the Hall-MHD shock structure problem by analyzing the above equation, it would be necessary to look for families of travelling wave solutions that become discontinuous as the viscosity and heat conductivity vanish. The disadvantage of this is that great effort will be expended on the global behavior of solutions to the system of travelling waves when the behavior we are trying to observe is local.
Instead we will use the method of matched asymptotic expansions to calculate travelling waves, an approach which will isolate the local behavior that is responsible for the shock structures.

The travelling wave problem is a singular perturbation problem since the small parameter multiplies the highest order terms
in the equations. When this parameter is equal to zero the order changes and fewer boundary conditions are required to determine the solution. These types of problems exhibit boundary layers and inner layers, which correspond to shocks. When the gradients of the dynamical variables are $O(1)$, the diffusive terms in Eqs. \ref{eq:twaves2} and \ref{eq:twaves4} are $O(\mu)$, so that to lowest order solutions of these equations mirror the behavior of the correspond inviscid solutions. On the other hand, in a region where there are steep gradients of the dynamical variables, the $\mu$ and $\kappa$ 
terms can be balanced by the derivative of ${\bf u}$ and $T_i$, which will be proportional to $1/\mu$ (we will assume $\kappa=O(\mu)$). In such a region ${\bf u}$ and $T_i$ will undergo a large change, but the magnetic field will
remain nearly constant, reflecting the predictions of the jump conditions written down in the previous section. The rest of the paper will
illustrate this heuristic picture in detail. 

The system Eq. \ref{eq:twaves1} through \ref{eq:twaves4} has two non-trivial distinguished limits, one represented by the variable $z$, 
and the other represented by the variable $\bar{z}=\frac{z}{\mu}$, with $z$ or $\bar{z}$ remaining finite as $\mu\rightarrow 0$. We distinguish between inner and outer solutions by appending a subscript,
${\bf B}_o$ represents an outer solution and ${\bf B}_i$ represents an inner solution. From here on out we will refer to the ion temperature $T_i$ with the variable $T$, so that the subscript $i$ stands for the inner solution only. The variable $B_z$ is constant in both the inner and outer solutions due to Eq. \ref{eq:BzConstant}.
The outer equations are defined by the outer variable $z$, and are the system \ref{eq:twaves1} to \ref{eq:twaves4}, but written in terms of an
outer solution:

\begin{subequations}\label{eq:OuterHall}
\begin{gather} \label{eq:HallOuterHydro}
\mu \de{{\bf u}_o}{z}=m {\bf u}_o +{\bf \hat{z}}\left(p_o+\frac{1}{2}B_o^2\right)- B_z {\bf B}_o - {\bf C}_{H} \\
-\frac{\epsilon B_z}{\rho_o}\calm\de{{\bf B}_{s,o}}{z}=u_{z,o} {\bf B}_{s,o} -B_z{\bf u}_{s,o} - {\bf C}_{B} \label{eq:HallOuterFlux} \\ 
\mu \frac{1}{2}\de{}{z} |{\bf u}_o|^2+\kappa\de{T_o}{z}=\frac{m}{2} |{\bf u}_o|^2 + \rho_o e_o u_{z,o}+ p_o u_{z,o} + {\bf B}_{s,o}\cdot{\bf C}_{B} -C_E \label{eq:HallOuterEnergy} 
\end{gather}\end{subequations}

The inner equations differ by a factor $\mu$ in the derivatives:

\begin{subequations}\label{eq:InnerHall}
\begin{gather} \label{eq:InneraHall}
\de{{\bf u}_i}{\bar{z}}=m {\bf u}_i +{\bf \hat{z}}\left(p_i+\frac{1}{2}B_i^2\right)- B_z {\bf B}_i - {\bf C}_{H} \\
-\frac{\epsilon B_z}{\mu\rho_i}\calm\de{{\bf B}_{s,i}}{\bar{z}}=u_{z,i} {\bf B}_{s,i} -B_z{\bf u}_{s,i} - {\bf C}_{B} \label{eq:HallInnerFlux} \\ 
\frac{1}{2}\de{}{\bar{z}} |{\bf u}_i|^2+k\de{T_i}{\bar{z}}=\frac{m}{2} |{\bf u}_i|^2 + \rho_i e_i u_{z,i}+ p_i u_{z,i} + {\bf B}_{s,i}\cdot{\bf C}_{B} -C_E \label{eq:HallInnerEnergy} 
\end{gather}\end{subequations}

The viscosity $\mu$ is a small parameter, and we will look for a solution in the 
form of an asymptotic series in $\mu$. In order to facilitate this we will make the assumption that $\kappa=k\mu$, where $k=O(1)$ and is constant.

\section{Solution of the Outer Equations to Lowest Order}

We expand the outer solution in an asymptotic series in $\mu$ of the form ${\bf U}_o(z)=\sum_{j=0}^{\infty}\mu^j{\bf U}^{(j)}_o(z)$ (where ${\bf U}_o$ is just being used to illustrate the general form of the series for each dynamical variable). Each equation becomes a series in powers of $\mu$, all orders of which must obey the equations. The lowest order equations in this hierarchy are the set of equations for travelling waves in inviscid Hall-MHD:

\begin{subequations}
\begin{gather} 
0=m {\bf u}^{(0)}_o +{\bf \hat{z}}\left(p^{(0)}_o+\frac{1}{2}B^{(0)2}_o\right)- B_z {\bf B}^{(0)}_o - {\bf C}_{H}\label{eq:OuterHydro} \\
0=\frac{m}{2} |{\bf u}^{(0)}_o|^2 + m e^{(0)}_o+ p^{(0)}_o u_{z,o}^{(0)} + {\bf B}_{s,o}^{(0)}\cdot{\bf C}_{B} -C_{E} \label{eq:OuterE}\\
\de{{\bf B}_{s,o}^{(0)}}{z}=\frac{m}{\epsilon B_z u_{z,o}^{(0)}}\calm\left(u_{z,o}^{(0)} {\bf B}_{s,o}^{(0)} -B_z{\bf u}_{s,o}^{(0)} - {\bf C}_{B}\right)  \label{eq:OuterB}
\end{gather}
\end{subequations}

All the integration constants, including $m$, are taken to be independent of the diffusivity $\mu$.
Eq. \ref{eq:OuterHydro} to \ref{eq:OuterB} form a system of differential-algebraic equations. We solve this system by first solving the algebraic equations, Eqs. \ref{eq:OuterHydro} and \ref{eq:OuterE}, to find an expression for the hydrodynamic variables in terms of ${\bf B}_{s,o}^{(0)}(z)$, i.e. ${\bf u}^{(0)}_{o}({\bf B}_{s,o}^{(0)})$ and $p^{(0)}_{o}({\bf B}_{s,o}^{(0)})$, and second using these expressions to close Eq. \ref{eq:OuterB} for the magnetic field ${\bf B}^{(0)}_o$. Eqs. \ref{eq:OuterHydro} and \ref{eq:OuterE} have been studied extensively in the development
of the theory of shock waves in compressible fluids as determining a shock's jump conditions. Under rather general conditions, for a given set of constants ${\bf C}_{B}$, 
$C_E$, ${\bf C}_H$, $m$, and $B_z$, and values of the magnetic field ${\bf B}^{(0)}_{s,o}(z)$, there are up to two solutions for the ${\bf u}_{o}$ and $p_o$. One condition that guarantees this property is that the functions $e$, $\tau=\frac{1}{\rho}$, $S$, and $p$, which define the thermodynamic properties of the plasma, satisfy the assumptions defining a \emph{Weyl fluid}\cite{Weyl}.
These are:
\begin{enumerate}
\item $\de{\tau}{p}\Big|_S < 0$
\item $\dnq{\tau}{p}{2}\Big|_S >0$
\item The pressure can take arbitrarily high values.
\item The thermodynamic state of the fluid is uniquely specified by $p$ and $\tau$, and the points $(p,\tau)$ representing possible states of the fluid form a convex region in the $(p,\tau)$ plane.
\end{enumerate}

A Weyl fluid must also satisfy the fundamental thermodynamic relation 
\begin{equation}
de=TdS-pd\tau. \label{eq:fundThermo}
\end{equation}
This class of fluids was introduced by Weyl as a general class of fluids for which shock waves of the type typically observed in gas dynamics arise. More commonly encountered types of fluids, such as ideal gasses or polytropic gasses, are special cases of Weyl fluids. 

Let us assume that the ion fluid is a Weyl fluid, that the equation of state of the ion fluid, $p_i=p_i(\tau,S)$ leads to
the satisfaction of the above conditions. 
Under this assumption we verify that the plasma, with its full equation of state, is also a Weyl Fluid. The internal energy of the plasma is the sum of the internal energy of the ions and electrons:
\begin{equation}
e=e_i-RT_e\log\frac{\tau}{\tau_0},
\end{equation}
satisfies Eq. \ref{eq:fundThermo}:
\begin{equation}
de=T_i dS_i-p_id\tau-R\frac{T_e}{\tau}d\tau
\end{equation} 
which satisfies Eq. \ref{eq:fundThermo} under the definition: $p=p_i+p_e=p_i+\frac{RT_e}{\tau}$. The entropy 
$S$ is equal to the ion entropy, $S=S_i$. Because the ion fluid is a Weyl fluid, $\de{p_i}{\tau}\big|_S<0$, which means $\de{p}{\tau}\big|_S=\de{p_i}{\tau}\big|_S-\frac{RT_e}{\tau^2}<0$, which implies $\de{\tau}{p}\big|_S<0$. Furthermore, $\dnq{p_i}{\tau}{2}\big|_S>0$ implies $\dnq{p}{\tau}{2}\big|_S=\dnq{p_i}{\tau}{2}\big|_S+\frac{2RT_e}{\tau^3}>0$. Combined with the fact that $\de{p}{\tau}\big|_S<0$ this implies that $\dnq{\tau}{p}{2}\big|_S>0$. The third condition
is clearly satisfied as $p=p_i+\frac{RT_e}{\tau}$ grows without bound when $\tau$ goes to zero. The combination of the third condition (the fact that $p_i$ can become arbitrarily high) and the fourth condition allow us to conclude that the set of allowable thermodynamic states for the ion fluid have the form $p_i>f(\tau)$ for $f$ convex and $\tau_{min}<\tau<\tau_{max}$. Therefore the allowable thermodynamic states of the full plasma have the form $p>f(\tau)+\frac{RT_e}{\tau}$, $\tau_{min}<\tau<\tau_{max}$, which is a convex
set due to the fact that $f(\tau)+\frac{RT_e}{\tau}$ is a convex function. Therefore the last condition is verified and the plasma 
satisfies the assumptions defining a Weyl Fluid.

Under the assumptions that the fluid is a Weyl fluid, the system comprised of Eqs. \ref{eq:OuterHydro} and \ref{eq:OuterE} has been well studied\cite{Courant}, and the properties of the solutions depend on the manner in which the problem is specified. If we specify the integration constants and the magnetic field ${\bf B}_s^{(0)}$, then there are either $0$, $1$, or $2$ solutions that correspond to physically realizable values of ${\bf u}$, $p$, and $\tau$. We characterize these solutions by Mach number $M=|u_z|/c_s$, where $c_s=\sqrt{\de{p}{\rho}\big|_S}$, which is possible. If $M>1$ the solution is a plus solution, written $\left({\bf u}^{(0)}_{>o}({\bf B}^{(0)}_{s,o}),p^{(0)}_{>o}({\bf B}^{(0)}_{s,o}),\tau^{(0)}_{>o}({\bf B}^{(0)}_{s,o})\right)$, if $M<1$ the solution is a minus solution, written $\left({\bf u}^{(0)}_{<o}({\bf B}^{(0)}_{s,o}),p^{(0)}_{<o}({\bf B}^{(0)}_{s,o}),\tau^{(0)}_{<o}({\bf B}^{(0)}_{s,o})\right)$. The functions $\left({\bf u}^{(0)}_{>o}({\bf B}^{(0)}_{s,o}),p^{(0)}_{>o}({\bf B}^{(0)}_{s,o}),\tau^{(0)}_{>o}({\bf B}^{(0)}_{s,o})\right)$ have as domain a subset $\calb_>$ of $\R^2$, consisting of all values of ${\bf B}^{(0)}_{s,o}$ for which there is a supersonic solution of Eq. \ref{eq:OuterHydro} and Eq. \ref{eq:OuterE}. The functions $\left({\bf u}^{(0)}_{<o}({\bf B}^{(0)}_{s,o}),p^{(0)}_{<o}({\bf B}^{(0)}_{s,o}),\tau^{(0)}_{<o}({\bf B}^{(0)}_{s,o})\right)$ have as domain a subset $\calb_<$, which consists of all values of ${\bf B}^{(0)}_{s,o}$ for which there is a subsonic solution of Eq. \ref{eq:OuterHydro} and Eq. \ref{eq:OuterE}. 

If instead of specifying the fluxes in Eqs. \ref{eq:OuterHydro} to \ref{eq:OuterB} $m$, ${\bf C}_B$, ${\bf C}_H$, $C_E$, and the magnetic field ${\bf B}_s^{(0)}$ we specify 
the magnetic field ${\bf B}_s^{(0)}$, the flux ${\bf C}_B$, the values of the hydrodynamic variables $\rho$, $p$, and 
${\bf u}$, whether the flow is supersonic or subsonic, and the normal velocity of the other solution ${\bf u}_{z,1}$ then there is another state with the same values of the constants and magnetic field, but on the opposite branch of solutions, so that if the original state was supersonic the second one would be subsonic and vice-versa. This is related to the specific shock structure problem that we will be intersted in studying: it means that if we specify the state ahead of the shock there will be a state behind the shock that satisfies the Rankine-Hugoniot conditions. 

The full set of differential algebraic equations reduce to the differential equations Eq. \ref{eq:OuterB} defined on  $\calb_>$ and $\calb_<$, using the solutions to Eqs. \ref{eq:OuterHydro} and \ref{eq:OuterE} to close the equations, 
i.e. on $\calb_>$ we write:
\begin{equation}
\de{{\bf B}_{s,o}^{(0)}}{z}=\frac{m}{\epsilon B_z u_{z,o>}^{(0)}({\bf B}^{(0)}_{s,o})}\calm\left(u_{z,o>}^{(0)}({\bf B}^{(0)}_{s,o}) {\bf B}_{s,o}^{(0)} -B_z{\bf u}_{s,o>}^{(0)}({\bf B}^{(0)}_{s,o}) - {\bf C}_{B}\right)
\end{equation}

 $\calb_>$ and $\calb_<$ correspond to surfaces in 
the full phase space of the system, which are defined as $\Xi_>=\calb_>\times(\left({\bf u}_>(\calb_>),p_>(\calb_>),\tau_>(\calb_>)\right)$ and $\Xi_<=\calb_<\times(\left({\bf u}_<(\calb_<),p_<(\calb_<),\tau_<(\calb_<)\right)$.

\begin{figure}
\begin{minipage}[t]{.45\linewidth}
\includegraphics[width=\linewidth]{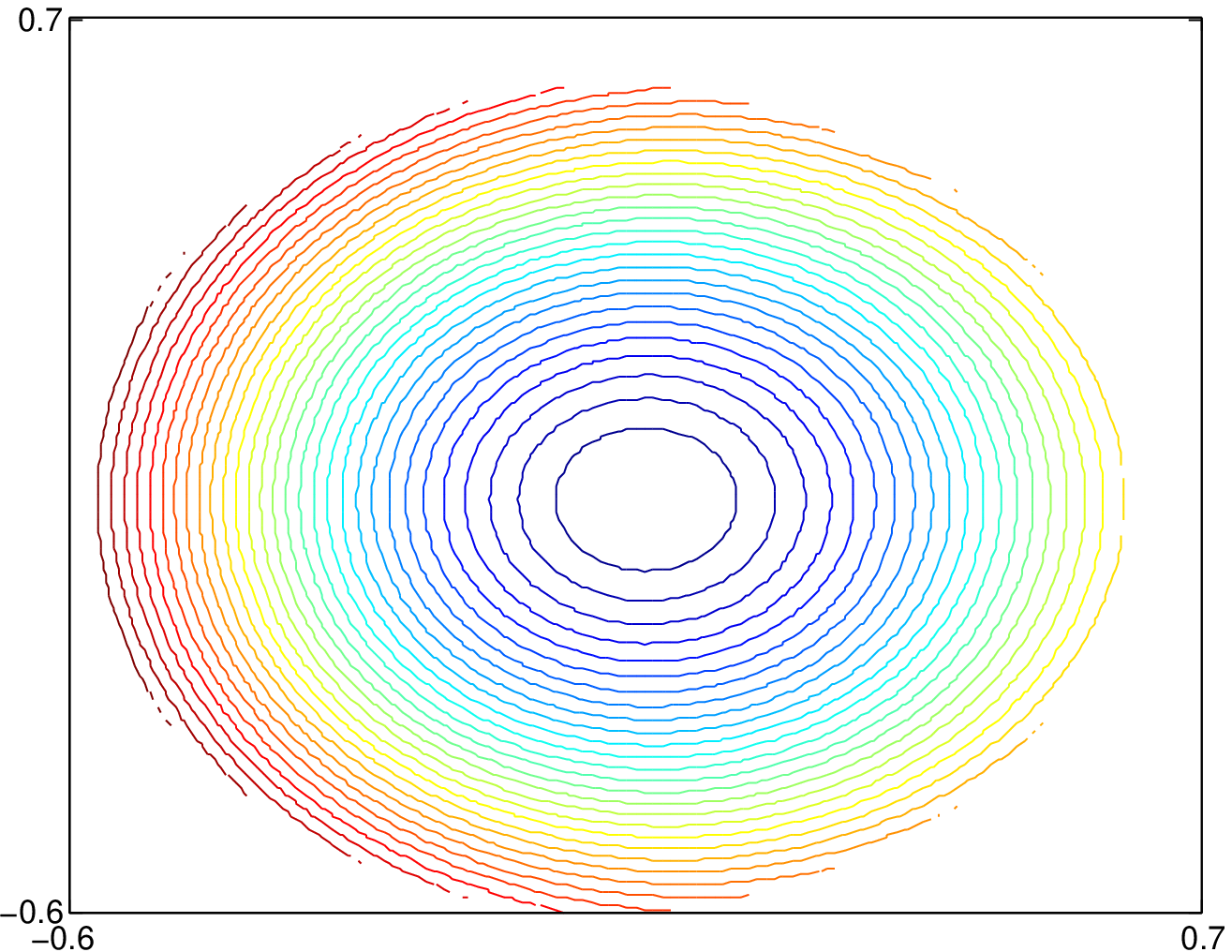}
\end{minipage}
\quad\quad
\begin{minipage}[t]{.45\linewidth}
\includegraphics[width=\linewidth]{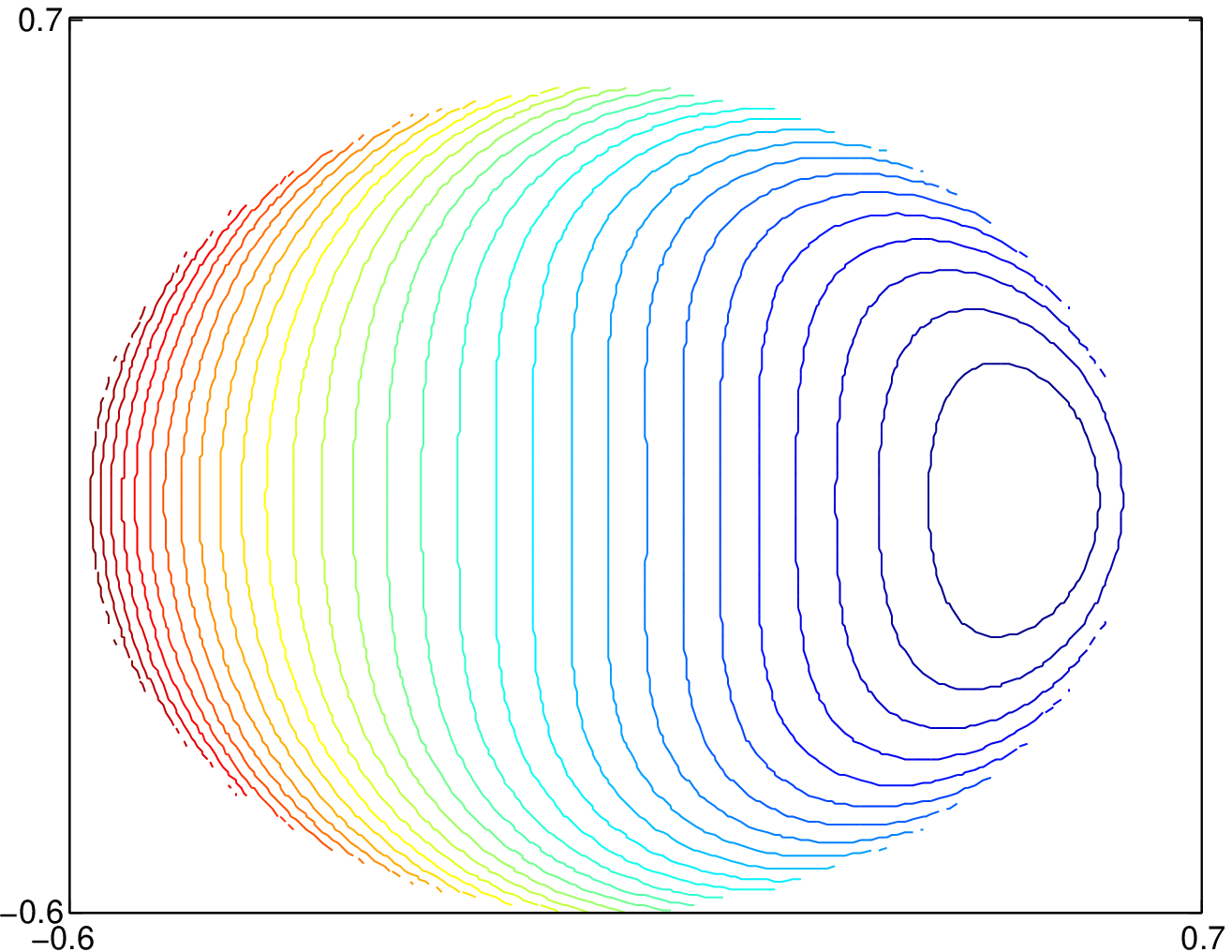}
\end{minipage}
\caption{Contour plots of the entropy $S^{(0)}_o$ on the outer domains $\Xi_>$ ($M>1)$ and $\Xi_<$ ($M<1)$, under the assumption that the ions are a polytropic fluid. The entropy is constant for continuous solutions of the outer equations, so that the level sets correspond to actual solutions. The boundary of the plot is the region $\Gamma$ where the Mach number is
1, at which point the equations become singular. There is an elliptic point near the center of both $\Xi_<$ and $\Xi_>$.}
\label{fig:orbits1}
\end{figure}

\begin{figure}
\begin{minipage}[t]{.45\linewidth}
\includegraphics[width=\linewidth]{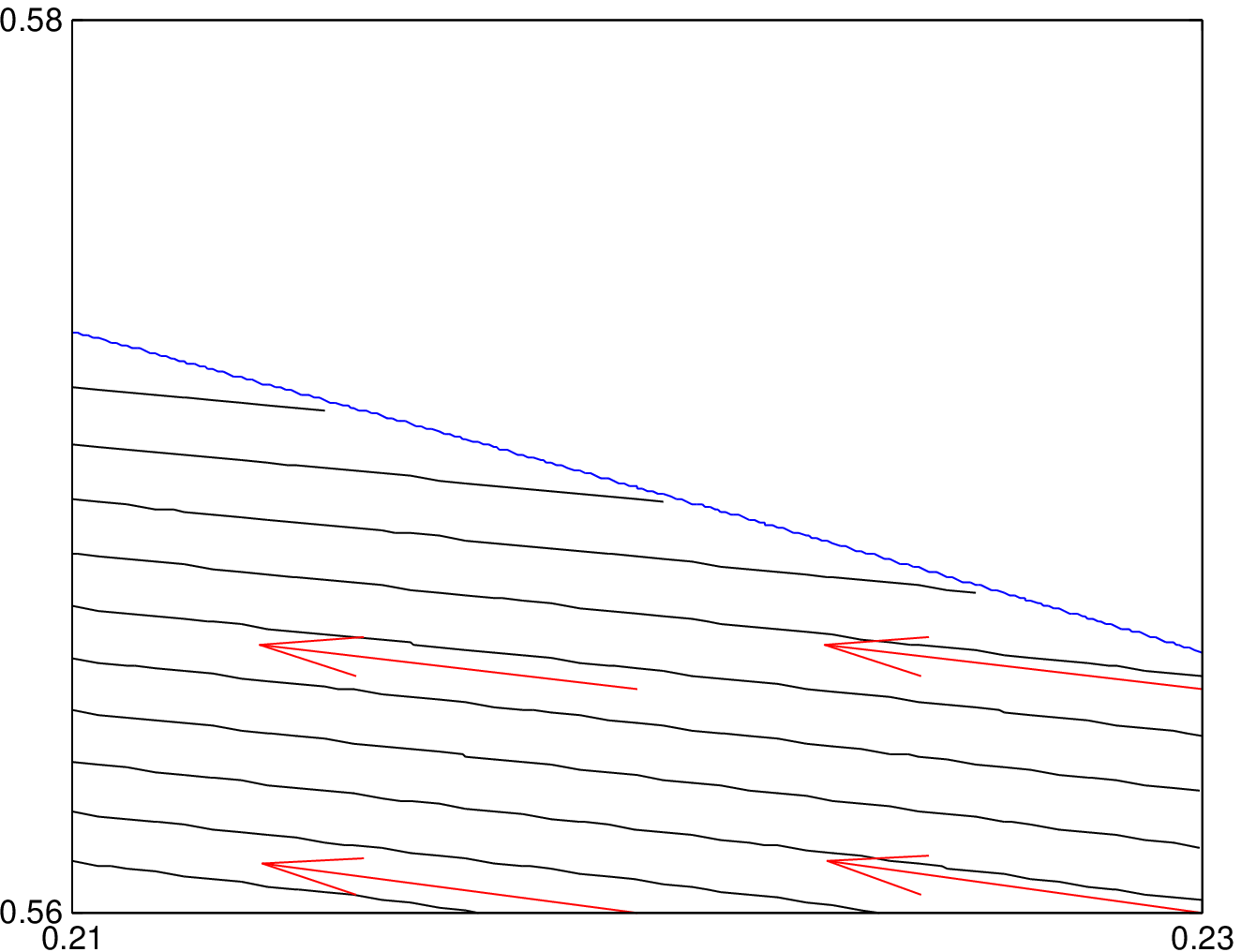}
\end{minipage}
\quad\quad
\begin{minipage}[t]{.45\linewidth}
\includegraphics[width=\linewidth]{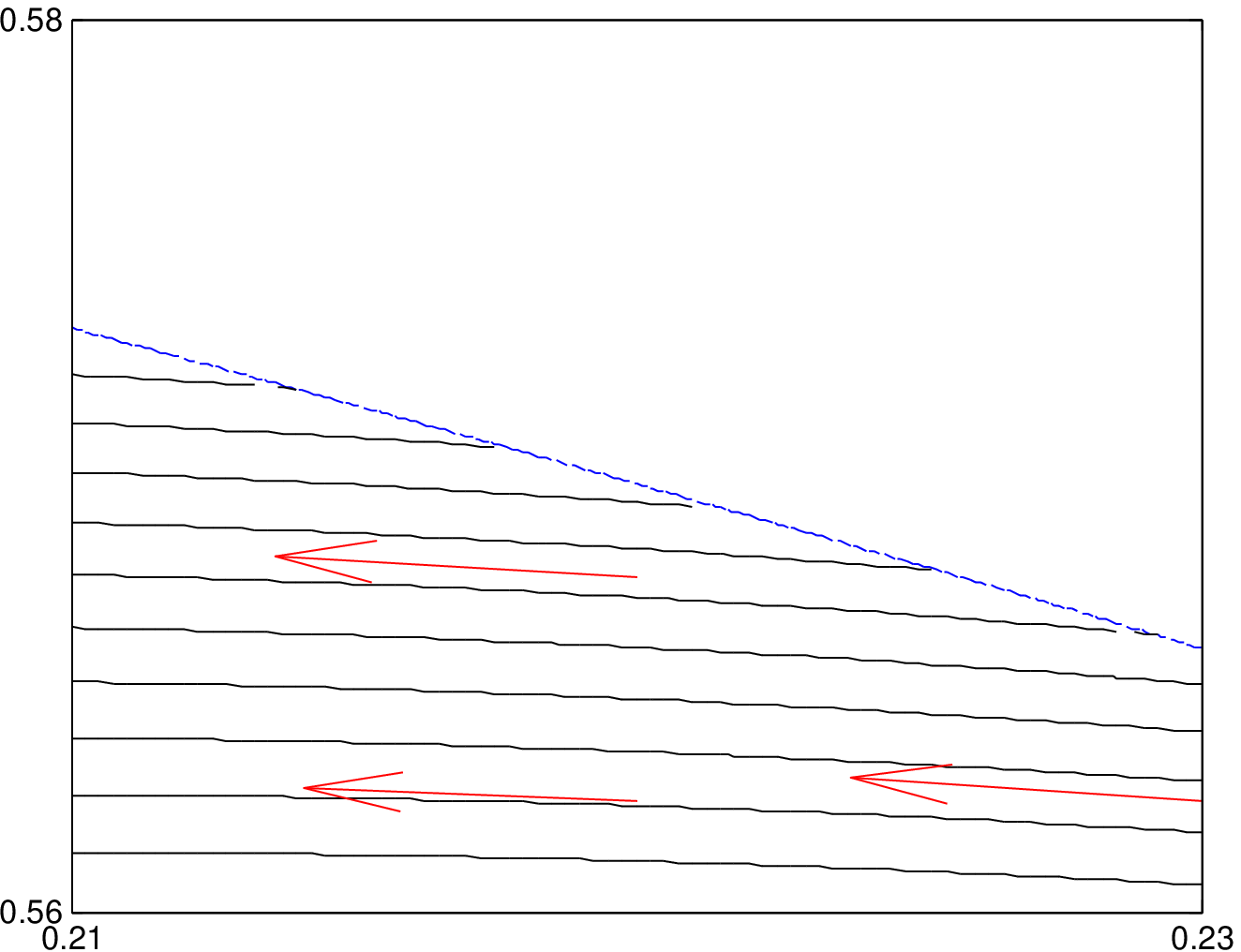}
\end{minipage}
\caption{Direction field and orbits in a neighborhood of $\Gamma$ on $\Xi_>$ and $\Xi_<$ respectively. When $M$ approaches
$1$ the vector fields become identical.}
\label{fig:CloseUp}
\end{figure}

\begin{figure}
\includegraphics[width=4in]{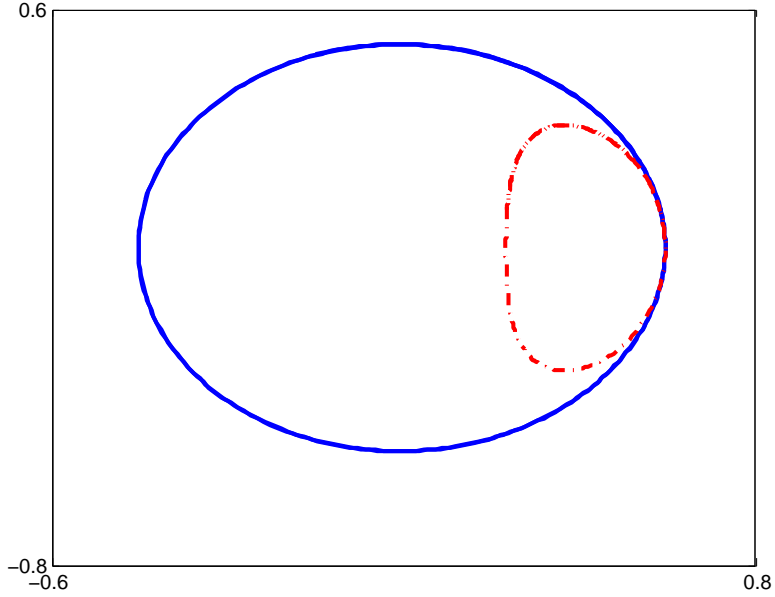}
\label{fig:tangentorbits}
\caption{Plot of orbits on $\Xi_>$ (solid) and $\Xi_<$ (broken) that are tangent to $\Gamma$.}
\end{figure}

Due to the abscence of viscosity and heat conductivity in Eq. \ref{eq:OuterHydro} to Eq. \ref{eq:OuterB}, the zeroth order entropy
$S^{(0)}_o$ is a constant to along the trajectory of the outer equations . The mass flux, momentum flux, and energy flux are each constant on the set $\Xi_>\cup\Xi_<$, but the entropy $S^{(0)}_o$ is not. Therefore the system of equations is integrable, and the orbits are the level sets of $S^{(0)}_o$ on $\Xi_>$ and $\Xi_<$. 
The orbits may be closed curves, but singular behavior arises when $M=1$, which occurs when $\Gamma=\bar{\Xi}_+\cap\bar{\Xi}_-$. The points on $\Gamma$ are also called \emph{impasse points}\cite{DAETravel,DAEAnalysis}. There are a number of different types of impasse points, the simplest of which are called either \emph{accessible} or \emph{inaccessible} simple singular points. These points are the points on $\Gamma$ where $\Gamma$ is not tangent to a level set of $S^{(0)}_o$ and which are not equilibrium points. Whether a simple singular point is accessible or inaccessible is determined by the direction of flow along the orbit through that point. At each inaccessible point of $\Gamma$ there are two solutions emerging, one going to $\Xi_>$ and one to $\Xi_<$, and at each accessible point there are two solutions, one coming from $\Xi_>$ and one
from $\Xi_<$, that reach $\Gamma$ and cannot be continued. The fact the solutions passing through a given simple singular point $\Gamma$ from $\Xi_>$ and $\Xi_<$ have the same behavior at $\Gamma$ is due to the fact that the right hand side of Eq. \ref{eq:OuterB} is a continuous function
of all of the dynamical variables, a fact which forces the vector fields on $\Xi_<$ and $\Xi_>$ to point in the same direction near $\Gamma$. 
The singular behavior of solutions at $\Gamma$ is typical of systems of differential algebraic equations, and it has been observed in the
similar context of resistive Hall-MHD\cite{HallDAE,HallIntermediate}. Almost every impasse point is of one of those two types, as there are only a finite set of equilibria of the outer equations (which have no particular reason to occur on $\Gamma$) and because in general the set $\Gamma$ is not a level set of $S^{(0)}$, so that generically the level sets of $S^{(0)}$ will cross it. Orbits near these singular points are plotted in Figure \ref{fig:CloseUp}. Because there will be both accessible and inaccessible singular points on $\Gamma$ there will generally be at least two points where orbits are tangent to $\Gamma$. In general there will be orbits on
both $\Xi_>$ or $\Xi_<$ that are tangent to these points, which means that it is possible to transition between on $\Xi_>$ and $\Xi_<$ on these special orbits. Two examples of such orbits are shown in Fig. \ref{fig:tangentorbits}. This singular behavior also implies that continuous travelling wave solutions of Hall-MHD can only change from subsonic to supersonic or vice-versa at special points; solutions that begin on $\Xi_>$ or $\Xi_<$ remain there or reach $\Gamma$. The behavior of trajectories near an equilibrium point that is embedded in $\Gamma$ is more complicated, in resistive Hall-MHD it is possible for solutions to cross between $\Xi_>$ and $\Xi_<$ at these points\cite{HallDAE}. Figure      \ref{fig:orbits1} shows level sets
of the entropy on $\Xi_>$ and $\Xi_<$ for an example of the outer equations.

\begin{centering}
\begin{figure}
\begin{minipage}[t]{.45\linewidth}
\includegraphics[width=\linewidth]{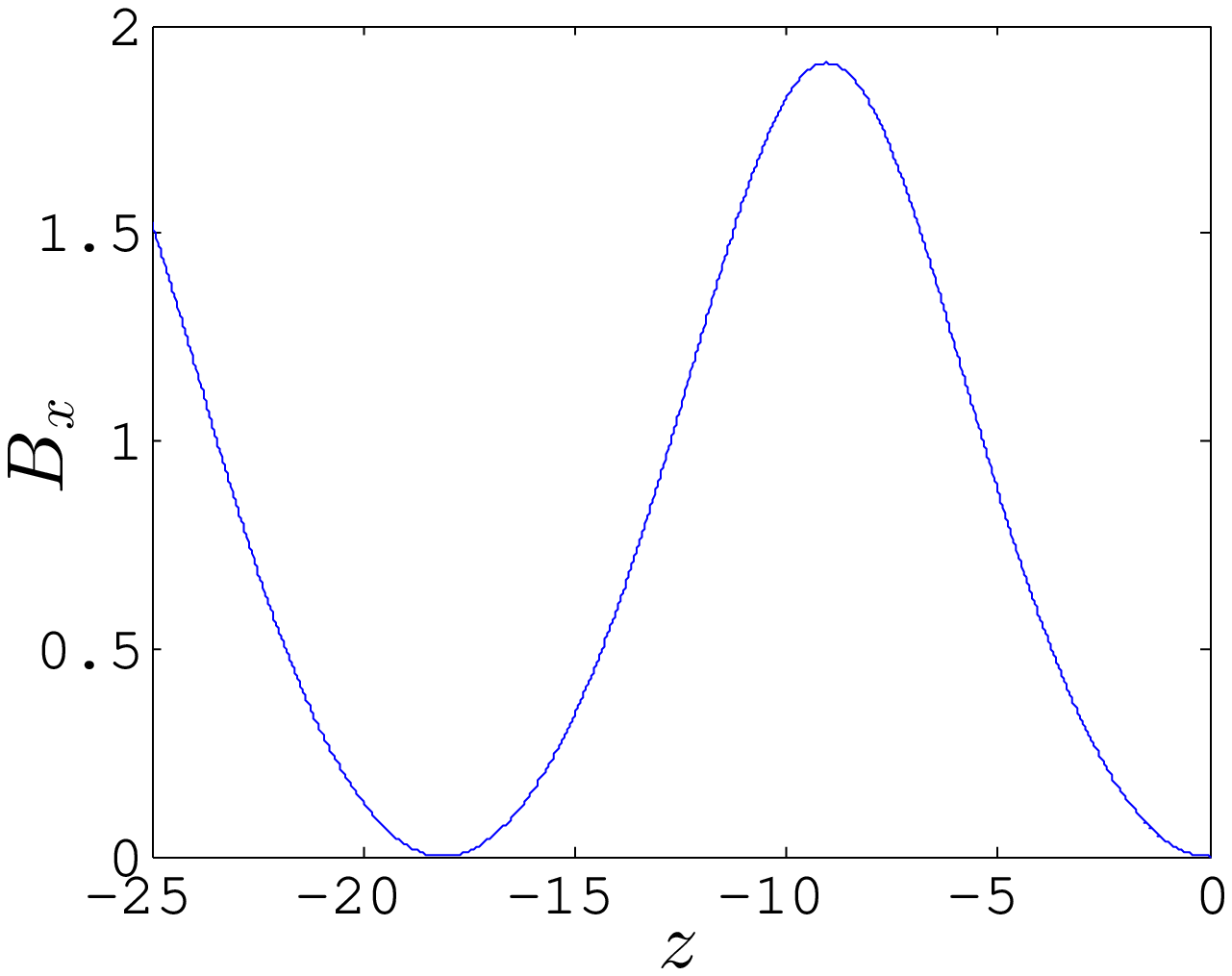}
\end{minipage}
\quad\quad
\begin{minipage}[t]{.45\linewidth}
\includegraphics[width=\linewidth]{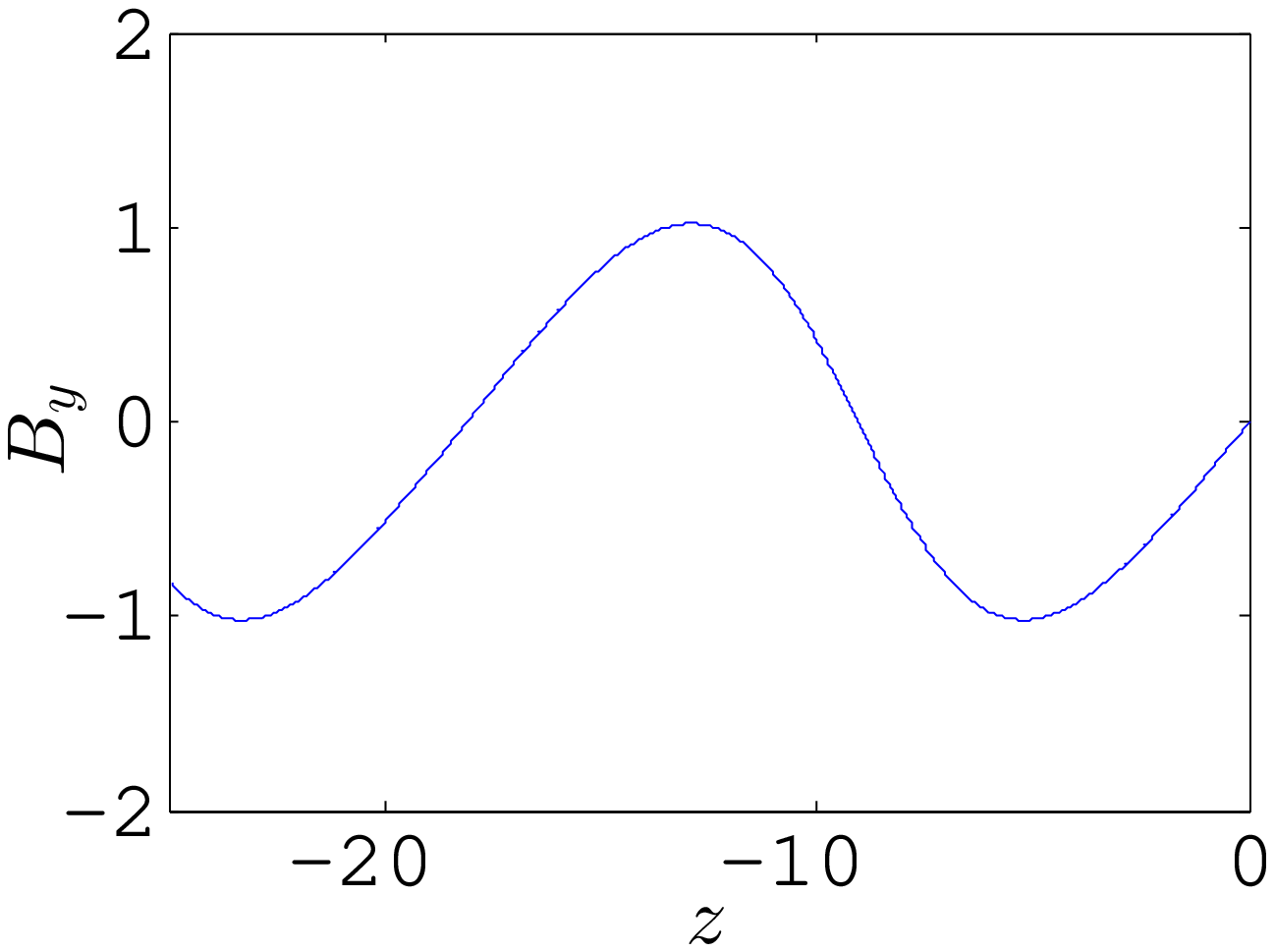}
\end{minipage}
\caption{Plots of the magnetic field components $B_x$ and $B_y$ of an outer solution on $\Xi_>$, with the assumption that the ions are a polytopic fluid.}
\label{fig:Bplus}
\end{figure}
\end{centering}

\begin{centering}
\begin{figure}
\begin{minipage}[t]{.45\linewidth}
\includegraphics[width=\linewidth]{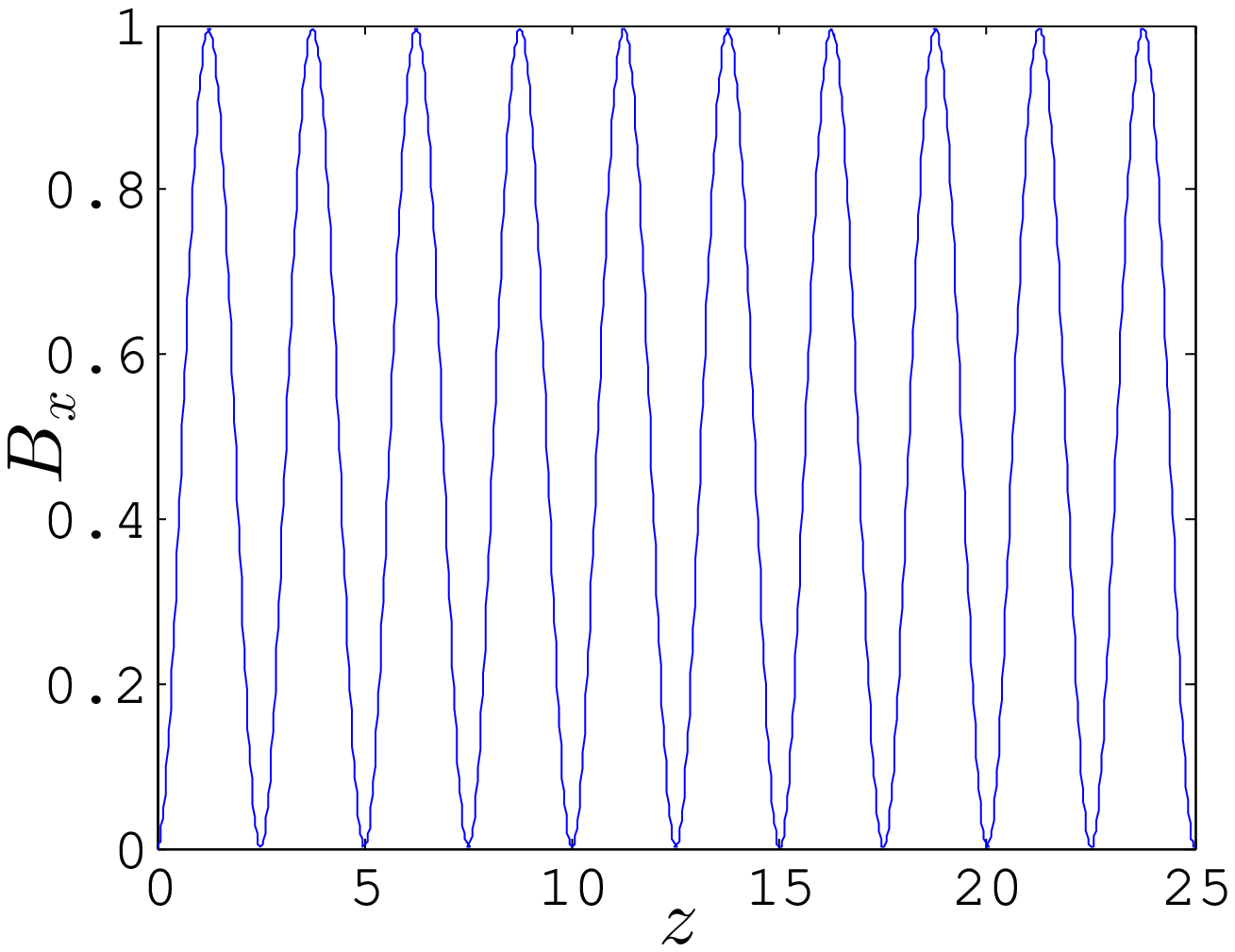}
\end{minipage}
\quad\quad
\begin{minipage}[t]{.45\linewidth}
\includegraphics[width=\linewidth]{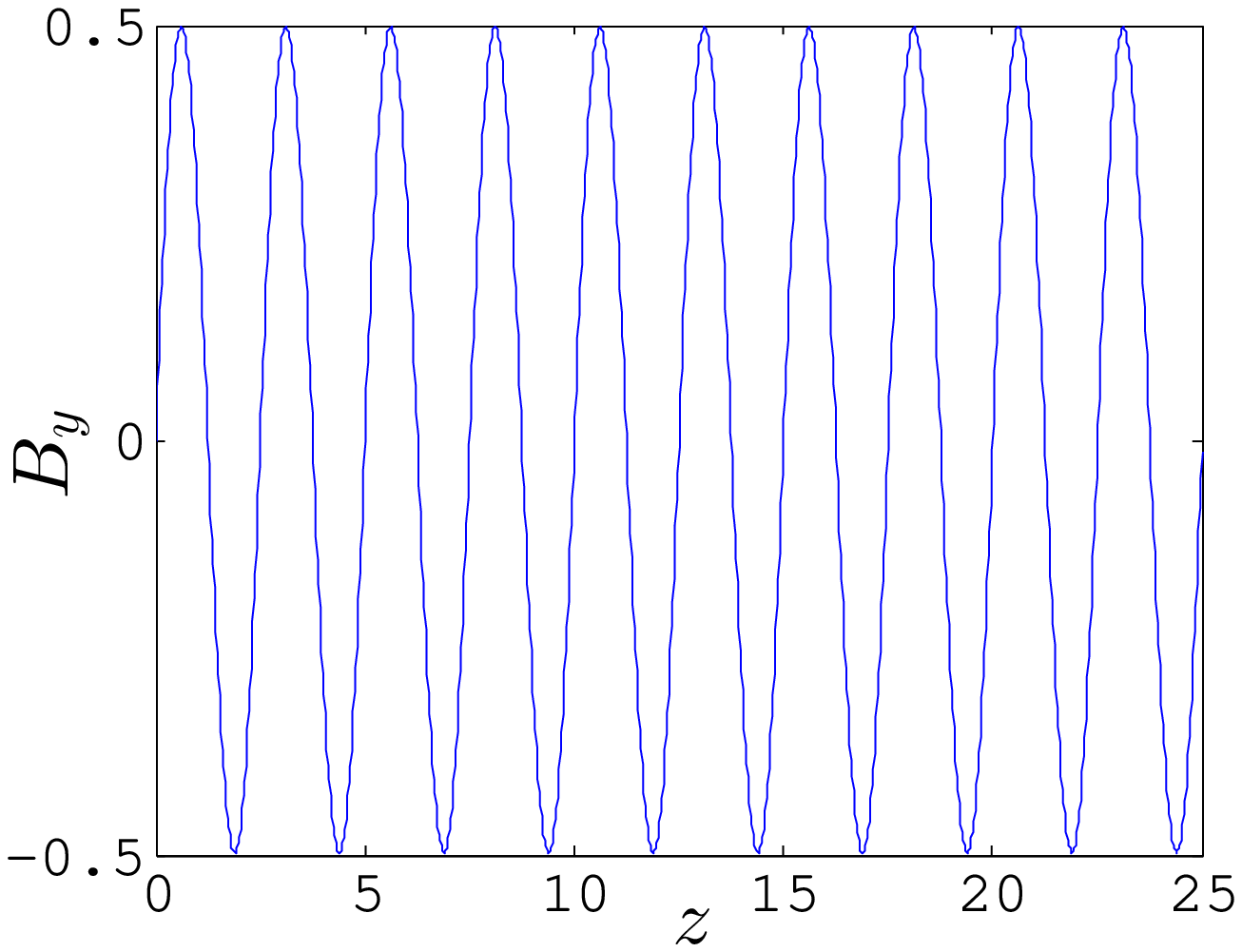}
\end{minipage}
\caption{Plots of the magnetic field components $B_x$ and $B_y$ of an outer solution on $\Xi_<$, with the assumption that the ions are a polytopic fluid.}
\label{fig:Bminus}
\end{figure}
\end{centering}

Given an initial value of $z$, the choice of an initial value of the magnetic field ${\bf B}^{(0)}_{s,o}$ and a branch, supersonic or subsonic, specifies a solution, which is unique and whose orbit is the level set of $S^{(0)}_{o\pm}$ passing through ${\bf B}^{(0)}_{s,o}$ as long as the initial ${\bf B}^{(0)}_s$ is in $\calb_+$ or $\calb_-$. An example solution for the special case of polytropic ions (which means that $A(S_i)=p_i\rho^{-\gamma}$, where $A(S_i)$ is a positive function), and computed using the fourth-order Runge-Kutta method with a spatial step of $h=.01$, is plotted in Figs. \ref{fig:Bplus} and \ref{fig:Bminus}.

\section{Solution of the Inner Equations to Lowest Order}
\label{sec:Inner}

The dynamical variables have steep gradients in the inner region, where it is assumed that the characteristic length scale is $O(\mu)$, so that the length scale with respect to $\bar{z}=\frac{z}{\mu}$ is $O(1)$.  We expand the inner solution in an asymptotic series in $\mu$, which takes the general form ${\bf U}_i(\bar{z})=\sum_{j=0}^{\infty}\mu^{j}{\bf U}^{(j)}_i(\bar{z})$. Each equation becomes a series in powers of $\mu$, each of which must be satsified. The lowest order equations in this hierarchy, which are derived from Eqs. \ref{eq:InneraHall} and \ref{eq:HallInnerEnergy}, are:

\begin{subequations}\label{eq:Inner}
\begin{gather} 
\de{{\bf u}^{(0)}_i}{\bar{z}}=m {\bf u}^{(0)}_i +{\bf \hat{z}}\left(p^{(0)}_i+\frac{1}{2}B^{(0)2}_i\right)- B_z {\bf B}^{(0)}_i - {\bf C}_{H}\label{eq:Innera} \\
\de{{\bf B}^{(0)}_{s,i}}{\bar{z}}=0 \label{eq:InnerFlux} \\ 
 \frac{1}{2}\de{}{\bar{z}} |{\bf u}^{(0)}_i|^2+k\de{T^{(0)}_i}{\bar{z}}=\frac{m}{2} |{\bf u}^{(0)}_i|^2 + \rho^{(0)}_i e^{(0)}_i u^{(0)}_{z,i}+ p^{(0)}_i u^{(0)}_{z,i} + {\bf B}^{(0)}_{s,i}\cdot{\bf C}_{B} -C_E\label{eq:InnerEnergy} 
\end{gather}\end{subequations}

Eq. \ref{eq:InnerFlux} implies that ${\bf B}^{(0)}_{s,i}$ is a constant function of $\bar{z}$. The magnetic field does not change to lowest order in the inner layer. 

Eqs. \ref{eq:Innera} and \ref{eq:InnerEnergy} reduce to the compressible Navier-Stokes equations of fluid dynamics, for the 
hydrodynamic variables. As such, the tangential components of velocity satisfy:
\begin{equation}
\de{{\bf u}_{s,i}^{(0)}}{\bar{z}}=m{\bf u}_{s,i}^{(0)}+const.
\end{equation}
The solution is:
\begin{equation}
{\bf u}_{s,i}^{(0)}={\bf A}e^{m\bar{z}}+{\bf D}
\end{equation}
where ${\bf A}$ and ${\bf D}$ are constants. We must pick ${\bf A}=0$ or ${\bf u}_{s,i}^{(0)}$ will diverge at either $\bar{z}=\infty$ or $-\infty$. Therefore ${\bf u}_{s,i}^{(0)}$ is constant.

The remaining system of two equations has been well-studied\cite{Gilbarg}, as it is the equation for viscous shock structures in the compressible Euler equation. There are up to two stationary points (independent of $\bar{z}$) with physical values of the hydrodynamic variables, which are solutions of the algebraic system:

\begin{subequations}\label{eq:InnerStationary}
\begin{gather} \label{eq:InneraStationary}
0=m {\bf u}^{(0)}_i +{\bf \hat{z}}\left(p^{(0)}_i+\frac{1}{2}B^{(0)2}_i\right)- B_z {\bf B}^{(0)}_i - {\bf C}_{H} \\
0=\frac{m}{2} |{\bf u}^{(0)}_i|^2 + \rho^{(0)}_i e^{(0)}_i u^{(0)}_{z,i}+ p^{(0)}_i u^{(0)}_{z,i} + {\bf B}^{(0)}_{s,i}\cdot{\bf C}_{B} -C_E \label{eq:InnerEnergyStationary} 
\end{gather}\end{subequations}
These two solutions \emph{are} presumably the limiting states of the inner solution as $\bar{z}\rightarrow\pm\infty$. 

\begin{centering}
\begin{figure}
\includegraphics[scale=.6]{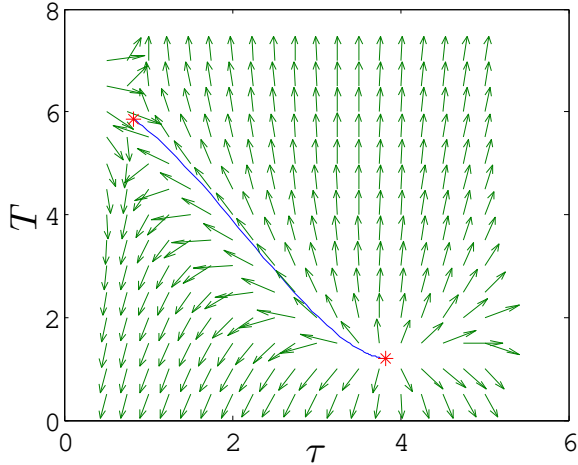}
\caption{Direction field for the inner equations written in terms of the variables $\tau$ and $T$. There are two stationary points, each represented by $*$, and a single heteroclinic connection between them.}
\label{fig:nsdirectionfield}
\end{figure}
\end{centering}

We assume that we are in the case where there are two stationary points. If that is the case, one of the stationary points has $M>1$ and is written $Z_+=({\bf u}^{(0)}_{i+}, p^{(0)}_{i+})$, and the other has $M<1$ and is written $Z_-=({\bf u}^{(0)}_{i-}, p^{(0)}_{-})$.
The point $Z_+$ is an unstable node and the point $Z_-$ is a saddle\cite{Gilbarg,Weyl}. Gilbarg\cite{Gilbarg} proved that there a unique heteroclinic orbit that connects the node at $\bar{z}=-\infty$ to the saddle at $\bar{z}=\infty$. Furthermore the rate of convergence to each saddle point is exponential in $\bar{z}$. This heteroclinic orbit is the inner
solution that we are interested in, so we define the inner solution $({\bf u}^{(0)}_{i}(\bar{z}),p^{(0)}_i(\bar{z}), {\bf B}^{(0)}_{s,i})$ so that the hydrodynamic part is the heteroclinic orbit constructed by Gilbarg. We provide in figure \ref{fig:nsdirectionfield} a plot of the direction field of the dynamical system corresponding to Eqs. \ref{eq:Innera} and \ref{eq:InnerEnergy}, including a plot of the heteroclinic connection between the two stationary points that corresponds to the desired inner solution. This solution was computed using a Chebyshev spectral method combined with Newton iterations, using the boundary condition that the solution be on the stable manifold of the saddle at $\bar{z}=\infty$ and a phase condition that half of the change in the dynamical variables across the interval occurs at the point $\bar{z}=0$.

Unlike in the outer equations, the entropy $S^{(0)}_i$ is not a constant in the inner layer. The value of
$S^{(0)}_i$ at the stationary point $Z_+$ is always greater than the value of $S^{(0)}_i$ at $Z_-$. Therefore $S^{(0)}_i(-\infty)<S^{(0)}_i(\infty)$, which reflects the fact that the entropy increases across the inner layer.

\section{Asymptotic Matching of the Outer and Inner Solutions}

Consider a pair of states of the fluid which satisfy the Rankine-Hugoniot conditions Eqs. \ref{eq:massflux1} through \ref{eq:Bcontinuity1}, and denote the $M>1$ state by $Z_-$ and the $M<1$ state by $Z_+$. Each
state determines a unique continuous solution of the non-diffusive ($\mu=0$) equations, and we can form an inviscid solution that contains a shock that satisfies the Rankine-Hugoniot conditions at the point $z=z_0$, which we take without loss of generality to be $z=0$, by forming piecewise solution equal to one of the continuous solutions
when $z>0$ and the other when $z<0$. 
The points $Z_-$ and $Z_+$ are stationary points of the inner equations, Eq. \ref{eq:InneraStationary} and Eq. \ref{eq:InnerEnergyStationary}. By the results of Section \ref{sec:Inner}, there is an inner solution $({\bf u}^{(0)}_{i}(\bar{z}),p^{(0)}_i(\bar{z}), {\bf B}^{(0)}_{s,i})$ whose limit at $\bar{z}=-\infty$ is the state $Z_-$ and whose limit at $\bar{z}=\infty$ is the state $Z_+$. 

The equations for the stationary points of the inner equations, Eqs. \ref{eq:InneraStationary} and \ref{eq:InnerEnergyStationary} are the same as the algebraic equations that define the hydrodynamic variables in terms of the magnetic field in the outer equations, Eq. \ref{eq:HallOuterHydro} and Eq. \ref{eq:HallOuterFlux}. Therefore the
stationary points of the inner equations lie on $\Xi_>$ and $\Xi_<$. Therefore the supersonic outer solution will match
with the supersonic stationary point at $z=0^-$ and $\bar{z}=-\infty$ and the subsonic outer solution at $z=0^+$ will match with the subsonic stationary point at $\bar{z}=\infty$.  

We can also check the value of the current of the inner and outer solution at $z=0$ and $\bar{z}=\pm\infty$. For the outer solution, the current ${\bf J}^{(0)}_{s,o}=\calm\de{B^{(0)}_{s,o}}{z}=\frac{m}{\epsilon B_z u_{z,o}^{(0)}}\calm\left(u_{z,o}^{(0)} {\bf B}_{s,o}^{(0)} -B_z{\bf u}_{s,o}^{(0)} - {\bf C}_{B}\right) $, which is from Eq. \ref{eq:OuterB}. In the inner region,
the $0$th order current is related to the derivative $1$st order magnetic field with respect to $\bar{z}$, ${\bf J}^{(0)}_{s,i}=\calm\de{B^{(1)}_{s,i}}{\bar{z}}$. The first order magnetic field is completely determined by the $0$th order solution, however, the 
formula for which is ${\bf J}^{(0)}_{s,i}=\calm\de{B^{(1)}_{s,i}}{z}=\frac{m}{\epsilon B_z u_{z,i}^{(0)}}\calm\left(u_{z,i}^{(0)} {\bf B}_{s,i}^{(0)} -B_z{\bf u}_{s,i}^{(0)} - {\bf C}_{B}\right)$. Therefore the equality of the outer and inner hydrodynamic variables also implies the equality of the outer and inner current.

We can form a uniformly valid composite solution by adding the inner and outer solutions and subtracting the matching part. Let the supersonic branch outer solution be written $({\bf u}^{(0)}_{o-}(z),p^{(0)}_{o-}(z),{\bf B}^{(0)}_{s,o-}(z))$, the subsonic branch outer solution be written $({\bf u}^{(0)}_{o+}(z),p^{(0)}_{o+}(z),{\bf B}^{(0)}_{s,o+}(z))$, and the inner solution be written
$({\bf u}^{(0)}_{i}(\bar{z}),p^{(0)}_{i}(\bar{z}),{\bf B}^{(0)}_{s,i}(\bar{z}))$. Let $H(z)$ be the Heaviside function.
Then the following composite solution is a uniformly valid solution:
\begin{subequations}
\begin{gather}
{\bf u}^{(0)}(z)=H(-z)\left({\bf u}^{(0)}_{o-}(z)-{\bf u}^{(0)}_{o-}(0)\right)+{\bf u}^{(0)}_{i}(\frac{z}{\mu})+H(z)\left({\bf u}^{(0)}_{o+}(z)-{\bf u}^{(0)}_{o+}(0)\right) \\
p^{(0)}=H(-z)\left(p^{(0)}_{o-}(z)-p^{(0)}_{o-}(0)\right)+p^{(0)}_{i}(\frac{z}{\mu})+H(z)\left(p^{(0)}_{o+}(z)-p^{(0)}_{o+}(0)\right) \\
{\bf B}^{(0)}_s=H(-z)\left({\bf B}^{(0)}_{s,o-}(z)-{\bf B}^{(0)}_{s,0-}(0)\right)+{\bf B}^{(0)}_{s,i}+H(z)\left({\bf B}^{(0)}_{s,o+}(z)-{\bf B}^{(0)}_{s,o+}(0)\right)
\end{gather}
\end{subequations}

In the limit
that $\mu$ goes to zero, this solution becomes a discontinuous travelling wave by virtue of the exponential convergence of the inner solution to a step function (in the variable $z=\mu\bar{z}$).
Across the discontinuity of the limiting solutions the values of the dynamical variables are equal to the value of
the inner solution at $\pm\infty=\bar{z}$. 

\begin{subequations}\begin{gather}
m[u_z]+[p] = 0 \label{eq:momentumflux11} \\
[{\bf u}_{s}] = 0 \label{eq:transversemomentum11}\\
\left[u_z{\bf B_{s}}+\frac{\epsilon B_n}{\rho}{\bf J_{s}}\right] = 0 \label{eq:magneticflux11} \\
\left[\left(\frac{1}{2}\rho {\bf u}^2+\rho e+p\right)u_z\right] = 0 \label{eq:energyflux11} \\
[{\bf B}]=0 \label{eq:Bcontinuity11}
\end{gather}\end{subequations}

Eqs. \ref{eq:momentumflux11} and \ref{eq:energyflux11}, which are the jump conditions for the hydrodynamic variables, are seen from substituting the limiting values $Z_-$ and $Z_+$ of
the outer equation into Eqs. \ref{eq:InneraStationary} and \ref{eq:InnerEnergyStationary} and subtracting the resulting expressions. Eq. \ref{eq:Bcontinuity11} comes from the constancy of the magnetic field in the inner solution, expressed by \ref{eq:InnerFlux}. 

To see Eq. \ref{eq:magneticflux11}, we substitute $Z_+$ and $Z_-$ into Eq. \ref{eq:OuterB} and subtract the two expressions, substituting $\calm \de{{\bf B}_s}{z}={\bf J}_o$. Another way to see that this relation holds would be to consider the next highest order equation for magnetic flux conservation in the inner region, which would give an expression for ${\bf J}^{(0)}_i=\calm\de{{\bf B}_{s,i}^{(1)}}{\bar{z}}$ in terms of the $0$th order inner variables. 

\begin{centering}
\begin{figure}
\begin{minipage}[t]{.45\linewidth}
\includegraphics[width=\linewidth]{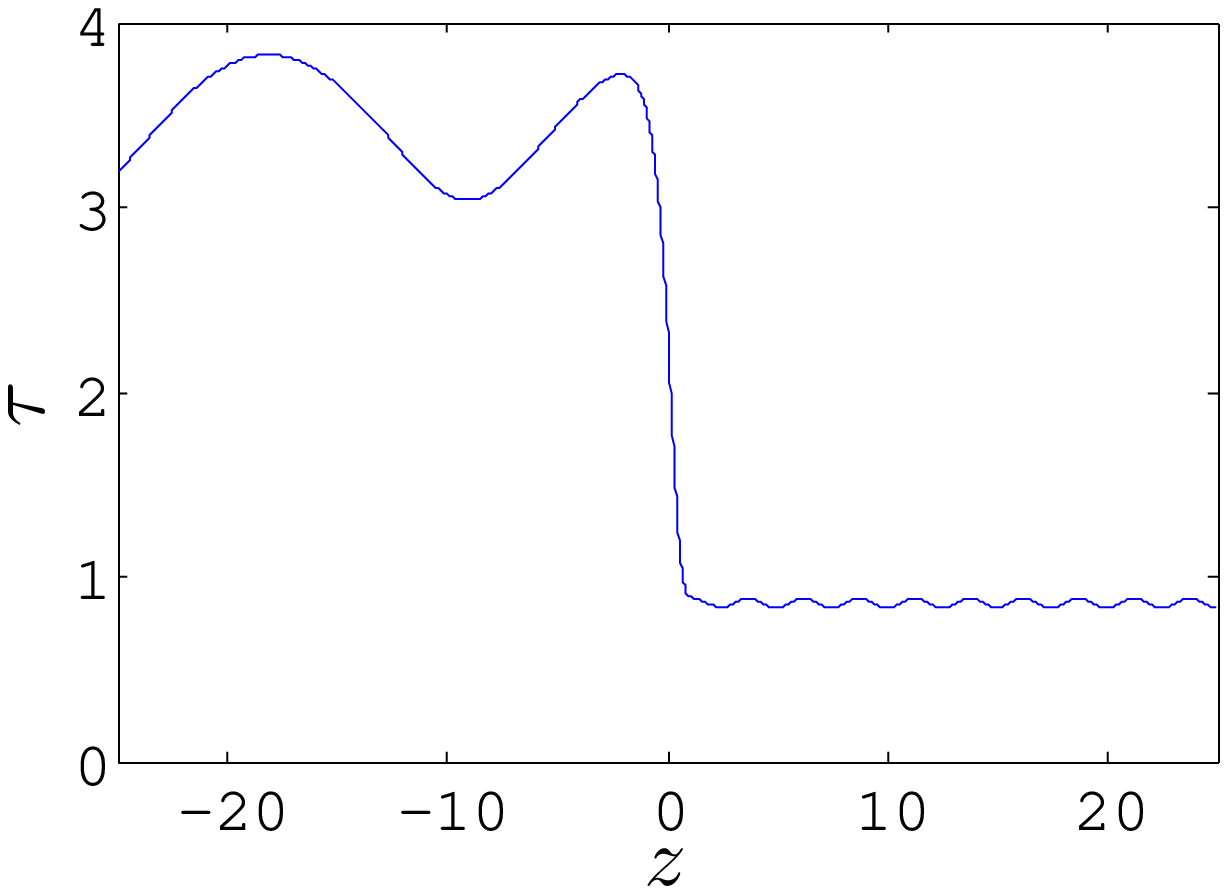}
\label{fig:vstruct}
\end{minipage}
\quad\quad
\begin{minipage}[t]{.45\linewidth}
\includegraphics[width=\linewidth]{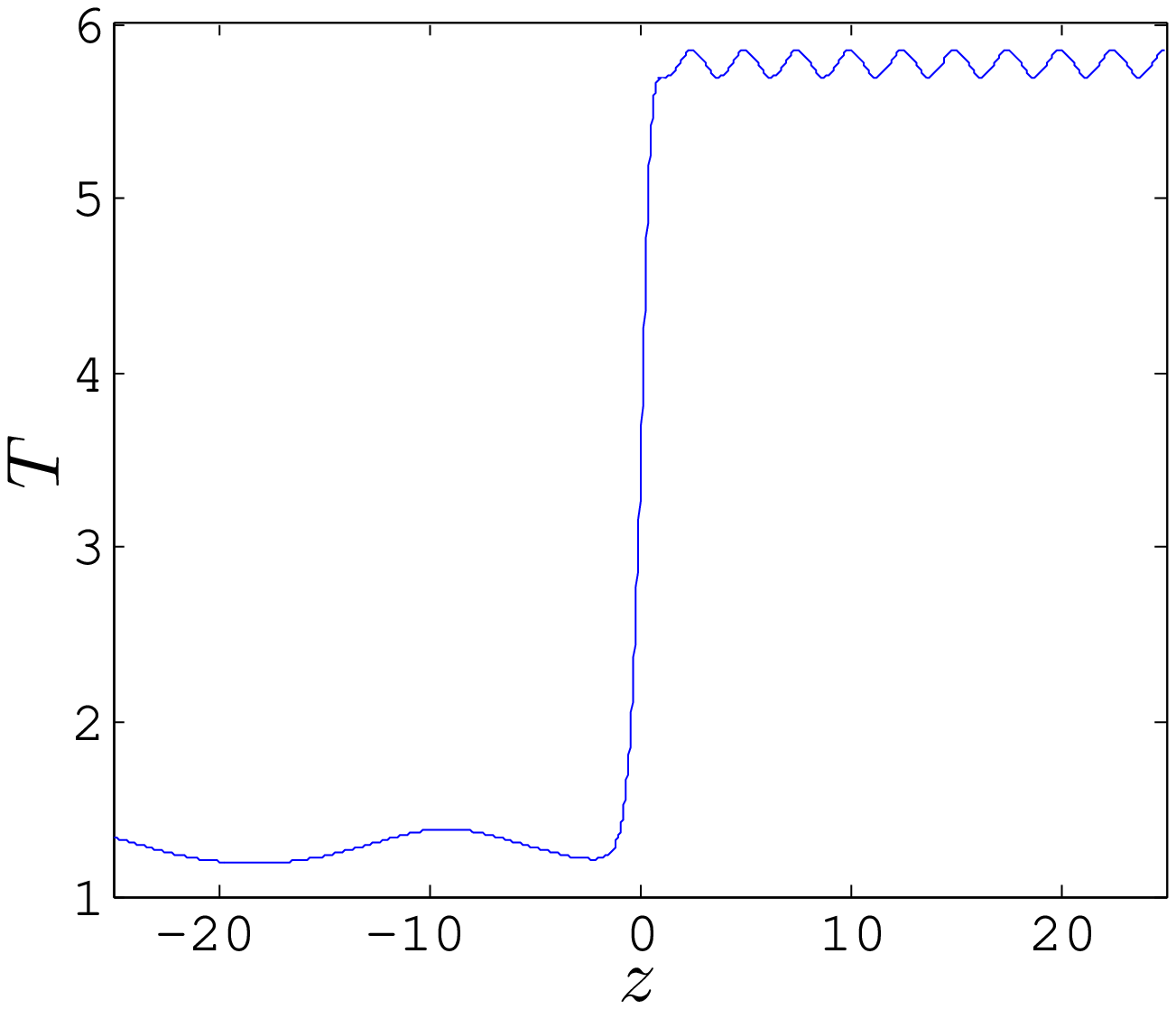}
\label{fig:tstruct}
\end{minipage}
\caption{Plot of $\tau$ and $T$ for a uniformly valid solution of the full equations in both the outer and inner regions.}
\end{figure}
\end{centering}

\begin{centering}
\begin{figure}
\includegraphics[width=.45\linewidth]{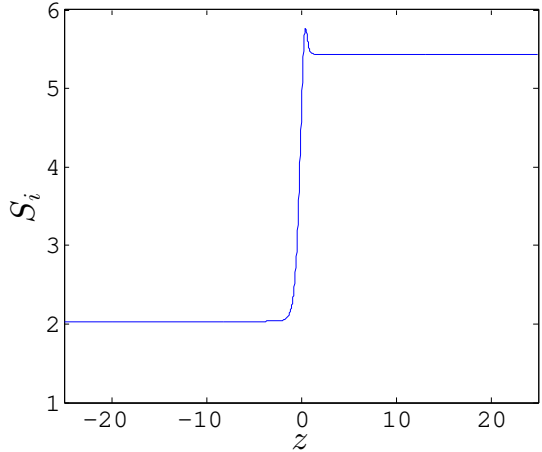}
\caption{Plot of $S$ for a uniformly valid solution of the full equations in both the outer and inner regions. The curve is not monotonic due to the
presence of heat conductivity.}
\label{fig:sstruct}
\end{figure}
\end{centering}

Therefore, for each discontinuous travelling wave solution of the Hall-MHD equations (with $\mu=0$) that also satisfies the above jump conditions and the entropy condition $m[S]>0$, there is a continuous composite viscous solution that realizes this discontinuity
in the limit that $\mu=0$. Therefore we conclude that the ion-acoustic shock wave has structure. In Figs. \ref{fig:vstruct} through \ref{fig:sstruct} we plot an example of a uniformly valid solution, computed under the assumption that the ion fluid is polytropic, demonstrating the sharp transition between oscillations near $\Xi_>$ and $\Xi_<$.

\begin{centering}
\begin{figure}
\includegraphics[width=.45\linewidth]{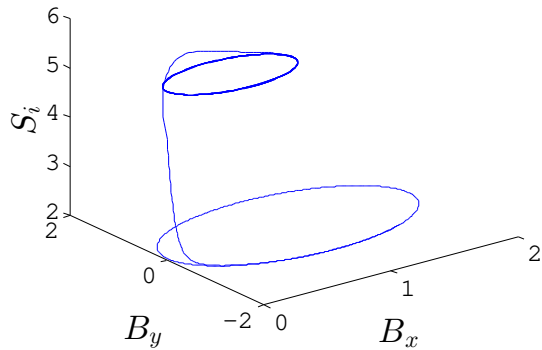}
\caption{Plot of a complete trajectory generated using a uniformly valid solution of the full equations, projected onto the three dynamical variables $B_x$, $B_y$, and $S_i$. Note the transition between oscillatory behavior on $\Xi_>$ to $\Xi_<$, on which the entropy remains roughly constant.}
\label{fig:trajectory}
\end{figure}
\end{centering}

\section{Conclusions}

We analyzed a family of solutions to the Rankine-Hugoniot equations in Hall-MHD. For any pair of states of the dynamical variables that satisfy these relations, we constructed a family of travelling wave solutions parametrized by the viscosity $\mu$ that converge to a piecewise continuous function with a discontinuity connecting the chosen pair of states. These solutions were constructed by the method of matched asymptotic expansions, combining an inner and outer solution that matched in an overlap region. The outer equations were a set of differential algebraic equations. Continuous solutions lay on a set of solutions of the algebraic part of the equations, either on a set with supersonic flow or on a set with subsonic flow. The orbits were level sets of the entropy. 

The inner equations were the compressible Navier-Stokes equations, which had two stationary points, one supersonic and one subsonic. All inner solutions had constant values of the magnetic field. There was a single solution connecting the supersonic stationary point to the subsonic. Each side of the inner solution match to a continuous outer solution, either supersonic or subsonic. 
The values on either side of the inner solution satisfy the Rankine-Hugoniot conditions and the entropy condition. In the limit of vanishing diffusivity the combined solution becomes discontinuous. From this we conclude that the ion-acoustic shocks have structure.

Shock waves and similar phenomenon are important in plasma physics because they are commonly occurring and because they arise in some of the most important problems. The theory of shock waves for Hall-MHD is interesting and important because of the equations are of a mixed type, they have the character or both hyperbolic and dispersive systems. This work continues the development of the theory of shock waves in HMHD that was by the development jump conditions for general shocks and the contact discontinuity that was begun by Hameiri\cite{HameiriHall}. Here we showed that the ion-acoustic shock has structure. The other shocks require more than one dimension, and the elucidation of the shock structure in those cases will be a topic of future work. We also intend to present a much more detailed solution, involving higher order terms in $\mu$ in a future manuscript.

This work is just part of a larger program to develop a better theoretical framework for Hall-MHD. This model is becoming increasingly important because of its role in the possible solution of some of the major problems of plasma physics, and because of its technological applications. The Hall effect might explain the growth rate and triggering of spontaneous reconnection in magnetospheric and tokamak plasmas, and also has applications to plasma switches and plasma thrusters. Further development of the mathematical basis of this model is warranted.

\acknowledgements{The authors gratefully acknowledge a number of discussions with Prof. Harold Weitzner. This work was supported by the US Department of Energy under grant no. DE-FG02-86ER53223.}

\end{document}